\documentclass[
  aip,
  jap,
  amsmath,amssymb,
  reprint,onecolumn,
  groupedaddress
]{revtex4-1}
\usepackage{graphicx,color}
\usepackage{amsmath}
\usepackage{natbib}
\usepackage{epsfig}
\begin{document}

\title{Practical formula for the shear viscosity of Yukawa fluids}

\author{Sergey Khrapak}
\affiliation{Institut f\"ur Materialphysik im Weltraum, Deutsches Zentrum f\"ur Luft- und Raumfahrt (DLR), 82234 We{\ss}ling, Germany; Joint Institute for High Temperatures, Russian Academy of Sciences, 125412 Moscow, Russia} 

\date{\today}

\begin{abstract}
A simple practical formula for the shear viscosity coefficient of Yukawa fluids is presented. This formula allows estimation of the shear viscosity in a very extended range of temperatures, from the melting point to $\simeq 100$ times the melting temperature. It demonstrates reasonable agreement with the available results from molecular dynamics simulations. Some aspects of the temperature dependence of the shear viscosity and diffusion coefficients on approaching the fluid-solid phase transition are discussed.                
\end{abstract}

\maketitle

\section{Introduction}

Studies of static and dynamical properties of Yukawa systems constitute an important interdisciplinary topic with
applications to strongly coupled plasmas, complex (dusty) plasmas, and colloidal suspensions. Significant efforts have been made over the years to understand the transport properties of such systems. In particular, this includes diffusion, viscosity, and thermal conductivity.
Extensive numerical simulations have been performed and a large amount of accurate data for the transport coefficients exist. What is often required in practical situations is a simple and accurate tool to estimate the transport coefficients in a broad range of parameters.     

The main purpose of this article is to present a useful practical expression to estimate the shear viscosity coefficient of three-dimensional Yukawa fluids in a wide regime of coupling and screening. The proposed formula is shown to describe quite well (deviations are within $\pm 10 \%$) the available results from numerical simulations in a very broad temperature range, from the melting temperature to $\simeq 100$ times the melting temperature. It can be particularly useful in the context of complex (dusty) plasmas and related soft weakly dissipative systems. 

\section{Background information}

The Yukawa systems are characterized by the repulsion between point-like charged particles immersed into neutralizing background, which provides screening. The corresponding interaction potential (also known as the Debye-H\"uckel or screened Coulomb potential) is
\begin{equation}\label{Yukawa}
\phi(r)=(Q^2/r)\exp(-r/\lambda),
\end{equation}
where $Q$ is the particle charge and $\lambda$ is the screening length. In the limit $\lambda\rightarrow\infty$ (i.e. screening is absent), the pure Coulomb interaction potential is recovered, corresponding to the one-component plasma (OCP) limit.~\cite{BrushJCP1966,BausPR1980} Yukawa potential is widely used as a reasonable first approximation for actual  interactions in three-dimensional isotropic complex plasmas and colloidal suspensions.~\cite{TsytovichUFN1997,FortovUFN,FortovPR,IvlevBook,KhrapakCPP2009,ChaudhuriSM2011,LampePoP2015} 
 
Yukawa systems are conventionally characterized by the two dimensionless parameters:~\cite{HamaguchiPRE1997} the coupling parameter $\Gamma=Q^2/aT$, and the screening parameter $\kappa=a/\lambda$, where $a=(4\pi n/3)^{-1/3}$ is the Wigner-Seitz radius and $n$ is the particle number density. The screening parameter $\kappa$ determines the softness of the interparticle interaction. It varies from the extremely soft and long-ranged Coulomb interaction at $\kappa\rightarrow 0$ to the hard-sphere-like interaction limit at $\kappa\rightarrow \infty$. In the context of complex plasmas and colloidal suspensions the relatively ``soft'' regime, $\kappa\sim {\mathcal O}(1)$, is of particular interest. Thermodynamics and dynamics of 3D Yukawa fluids and crystals in this regime have been extensively studied in the literature.~\cite{HamaguchiJCP1996,HamaguchiPRE1997,RobbinsJCP1988,ToliasPRE2014,KhrapakPRE2014,KhrapakPRE02_2015,KhrapakJCP2015,
YurchenkoJPCM2016,KhrapakPPCF2015}

Depending on the values of $\Gamma$ and $\kappa$, Yukawa systems can form either a fluid or, at sufficiently high $\Gamma$, a solid phase. There is no gas-liquid phase transition, because attraction is absent. In the solid phase the body-centered-cubic (bcc) or face-centered-cubic (fcc) lattices can be thermodynamically stable (bcc lattice is thermodynamically favorable at weak screening). The values of the coupling parameter at which the fluid-solid phase transition occurs are usually denoted $\Gamma_{\rm m}$.  The dependence $\Gamma_{\rm m}(\kappa)$ obtained from molecular dynamics (MD) simulations is available.~\cite{HamaguchiJCP1996,HamaguchiPRE1997} A simple formula~\cite{VaulinaJETP2000,VaulinaPRE2002}
\begin{equation}\label{melt}
\Gamma_{\rm m}(\kappa)\simeq \frac{172 \exp(\alpha\kappa)}{1+\alpha\kappa+\tfrac{1}{2}\alpha^2\kappa^2},
\end{equation}
is in a rather good agreement with the tabulated data in the regime $\kappa\lesssim 5$.  Here the constant $\alpha=(4\pi/3)^{1/3}\simeq 1.612$  is the ratio of the mean interparticle distance $\Delta=n^{-1/3}$ to the Wigner-Seitz radius $a$. 
The occurrence and location of a glass phase on the phase diagram of Yukawa systems have been investigated.~\cite{SciortinoPRL2004,YazdiPRE2014} This regime is not considered below, the attention is focused on the strongly coupled fluid phase.           

\section{Shear viscosity of Yukawa fluids}

\subsection{Numerical data}
 
Shear viscosity of single-component 3D Yukawa fluids has been extensively studied using MD simulations,~\cite{VieillefossePRA1975,SanbonmatsuPRL2001,SaigoPoP2002,SalinPRL2002,SalinPoP2003,DonkoPRE2008,DonkoPRE2010,MithenCPP2012} see in particular an overview in Ref.~\onlinecite{DaligaultPRE2014}. Despite the apparent simplicity of the single-component Yukawa model, accurate determination of the shear viscosity by MD simulation is not trivial. This can be exemplified by the significant discrepancies in the results obtained over the years by different authors.~\cite{DaligaultPRE2014} In the following, the two data sets tabulated in Refs.~\onlinecite{DonkoPRE2008} and \onlinecite{DaligaultPRE2014} are chosen as the most accurate results presently available. 

\begin{figure}
\includegraphics[width=10cm]{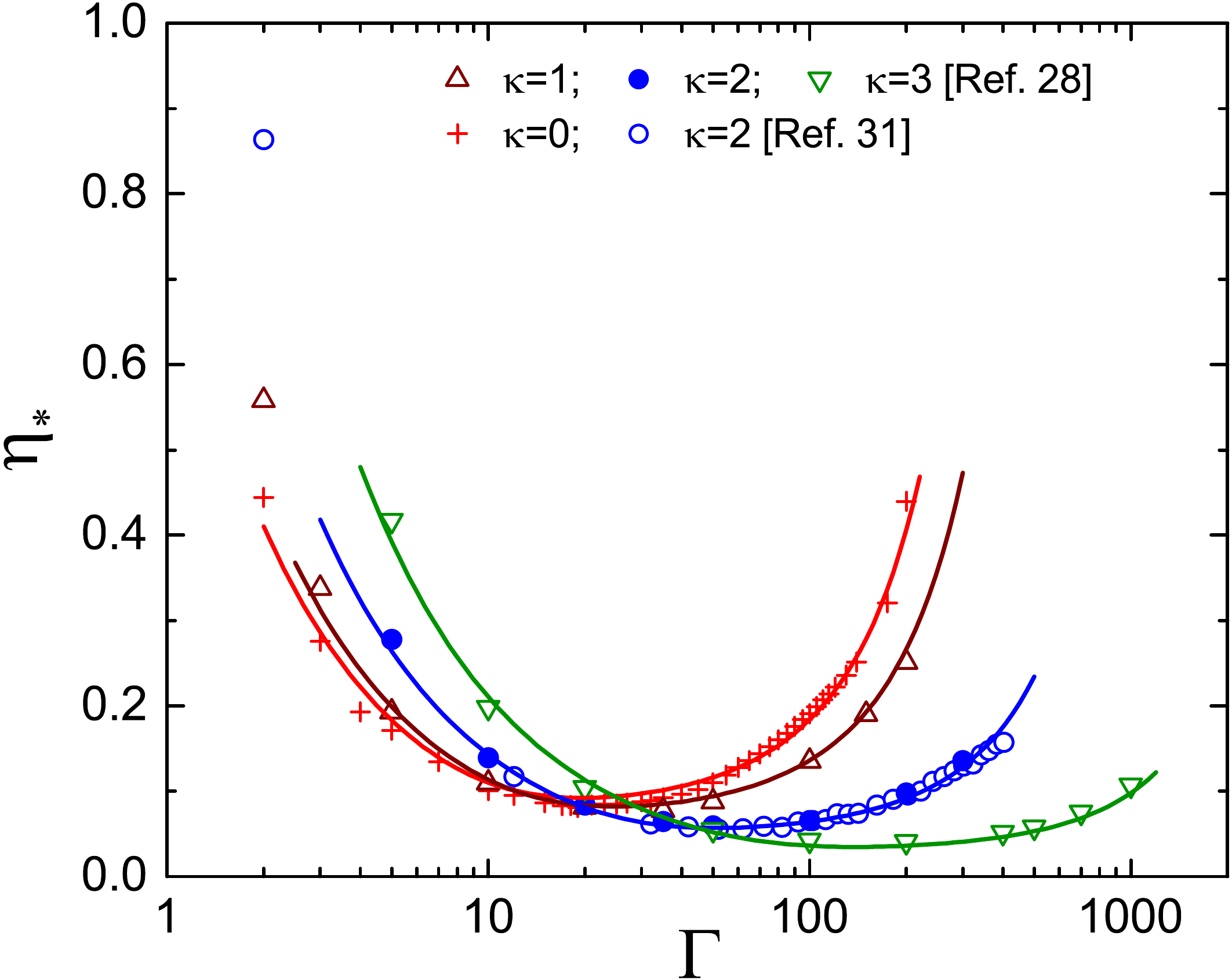}
\caption{Reduced shear viscosity coefficient of Yukawa fluids $\eta_*$,  as a function of the coupling parameter $\Gamma$. The symbols correspond to the data tabulated in Refs.~\onlinecite{DonkoPRE2008,DaligaultPRE2014}. The solid curves are calculated with the help of a practical interpolation formula (\ref{inter}) suggested in this work. Several points corresponding to the very weakly coupled gaseous regime have been excluded from the analysis.}
\label{Fig1}
\end{figure}

The original data from Refs.~\onlinecite{DonkoPRE2008,DaligaultPRE2014} used in this work are plotted in Fig.~\ref{Fig1}. Here the conventional (for plasma physics) normalization is used: The reduced shear viscosity coefficient is     
\begin{equation}\label{Norm1}
\eta_*=\frac{\eta}{m n a^2 \omega_{\rm p}},
\end{equation}
 where $m$ is the particle mass and $\omega_{\rm p}=\sqrt{4\pi Q^2 n/m}$ is the plasma frequency. In Fig.~\ref{Fig1}, $\eta_*$ is plotted versus the coupling parameter; symbols are the tabulated MD data points~\cite{DonkoPRE2008,DaligaultPRE2014} and the curves correspond to a practical interpolation formula derived later in this work. 
 
\subsection{Normalization} 
 
Historically, the use of the plasma frequency $\omega_{\rm p}$ in presenting reduced transport coefficients of OCP and Yukawa systems [like in Eq.~(\ref{Norm1})] originates from the pioneering works of Hansen and collaborators.~\cite{HansenPRA1975,VieillefossePRA1975} More general macroscopic reduction parameters, not limited to plasma physics context, have been suggested by Rosenfeld.~\cite{RosenfeldPRA1977,RosenfeldJPCM2001} Namely, the mean interparticle separation $\Delta = n^{-1/3}$ and the thermal velocity $v_{\rm T}=\sqrt{T/m}$ have been used as the units of length and velocity. In these units the reduced shear viscosity coefficient becomes
\begin{equation}
\eta_{\rm R} = \frac{\eta}{m v_{\rm T} n^{2/3}}.
\end{equation}        
This form of the reduced viscosity coefficient is suggested by an elementary kinetic theory formula for viscosity,  $\eta\sim n m \ell v_{\rm T}$, for a dense medium of particles with thermal velocities $v_{\rm T}$ and a mean free path between collisions $\ell$, which is of the order of the average interparticle distance.~\cite{RosenfeldJPCM2001} The relation between the two reduced viscosity coefficients is $\eta_{\rm R}=\eta_*(4\pi /3)^{-5/6}\sqrt{4\pi \Gamma}$.

Rosenfeld used this normalization first to suggest a quasi-universal {\it excess-entropy} scaling of the properly reduced atomic transport coefficients of simple fluids.~\cite{RosenfeldPRA1977,RosenfeldJPCM1999} Then he came  to a conclusion that for sufficiently soft interaction potentials the excess-entropy scaling should result in a quasi-universal {\it freezing-temperature} scaling of the reduced transport coefficients (simply because the excess entropy itself scales quasi-universally with the temperature reduced by its value at freezing).~\cite{RosenfeldJPCM2001,RosenfeldPRE2000} Various other arguments have been put forward in favor of the freezing-temperature scaling of the shear viscosity.~\cite{SaigoPoP2002,VaulinaJETP2004,VaulinaAIP2005,Kaptay2005,MuratovJETPLett2007,DingSM2015,CostigliolaJCP2018}  Perhaps the most solid basis behind the freezing-temperature scaling is the isomorph theory developed recently.~\cite{GnanJCP2009,DyreJPCB2014}    
With isomorphs defined as the lines of constant excess entropy in the thermodynamic phase diagram, the reduced viscosity is
constant along an isomorph because the properly reduced dynamics is.~\cite{CostigliolaJCP2018} The freezing curve is an approximate liquid-state isomorph. Parallel curves, characterized by fixed ratios $T_{\rm m}/T\equiv \Gamma/\Gamma_{\rm m}$  should be approximate isomorphs, too. Application of the isomorph theory to Yukawa systems has been recently discussed in detail.~\cite{VeldhorstPoP2015}

\begin{figure*}
\includegraphics[width=14cm]{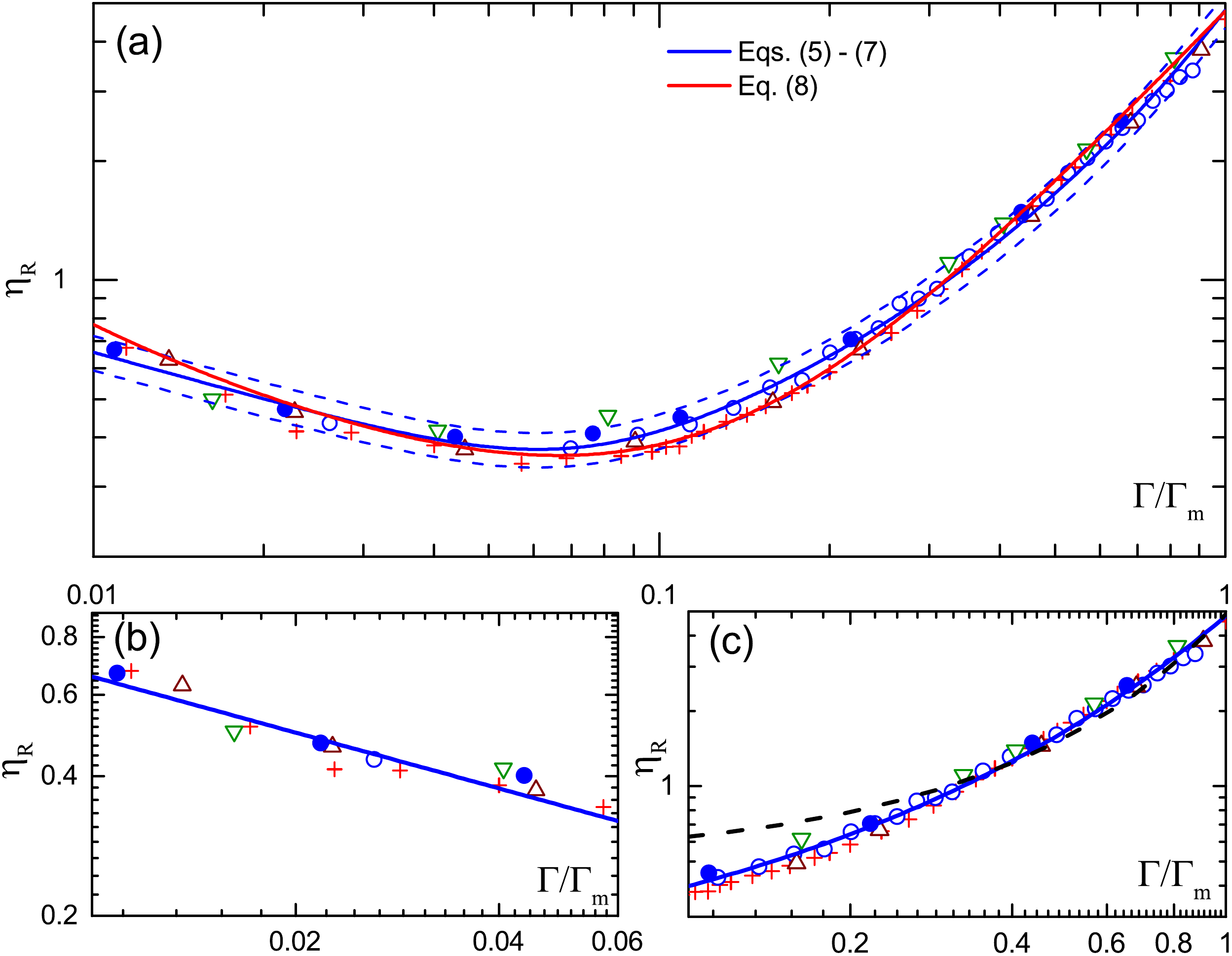}
\caption{Reduced coefficient of shear viscosity $\eta_{\rm R}$ of Yukawa fluids versus the reduced coupling parameter $\Gamma/\Gamma_{\rm m}$. Symbols correspond to the data tabulated in Refs.~\onlinecite{DonkoPRE2008,DaligaultPRE2014} (the notation is the same as in Fig.~\ref{Fig1}). In (a) the entire range of coupling considered, $0.01\leq \Gamma/\Gamma_{\rm m}\leq 1$, is shown.
In addition, the interpolation formula (\ref{inter}) is plotted as a blue solid curve. The dashed curves mark $\pm 10\%$ deviation region. The red curve is the modified OCP fit of Eq.~(\ref{OCPfit}). In (b) the data corresponding to the moderately coupled regime ($0.01\lesssim \Gamma/\Gamma_{\rm m}\lesssim 0.06$) are shown, the solid line corresponds to the fit of Eq.~(\ref{fit1}). In (c) the data corresponding to the strongly coupled regime ($0.1\lesssim \Gamma/\Gamma_{\rm m}\lesssim 1$) are shown, the solid line corresponds to the fit of Eq.~(\ref{fit2}). The dashed line is the exponential fit of the form $\eta_{\rm R}\propto \exp(2.28\Gamma/\Gamma_{\rm m})$.  
}
\label{Fig2}
\end{figure*}

Having all this in mind, the original data from Refs.~\onlinecite{DonkoPRE2008,DaligaultPRE2014} plotted in Fig.~\ref{Fig1} have been re-scaled  to produce the dependence $\eta_{\rm R}$ on $\Gamma/\Gamma_{\rm m}$. The results are shown in Fig.~\ref{Fig2} (a). Here the data corresponding to the fluid regime with $0.01\leq \Gamma/\Gamma_{\rm m}\leq 1$ are plotted (a few points corresponding to the very weakly coupled gaseous regime were not retained for further analysis). In contrast to Fig.~\ref{Fig1}, the data for $\eta_{\rm R}$ versus $\Gamma/\Gamma_{\rm m}$ appear to collapse onto a single universal curve. While some scattering of the data points is still present, no clear systematic dependence on the screening parameter $\kappa$ is evident. The scattering is probably related to an uncertainty in evaluating shear viscosity from MD simulations. The quality of the collapse improves on approaching the fluid-solid (freezing) phase transition. Below we propose a practical expression of the form $\eta_{\rm R}={\mathcal F}(x)$, where  $x=\Gamma/\Gamma_{\rm m}=T_{\rm m}/T$, which provides a good estimate of the shear viscosity of Yukawa fluids in a wide region of the phase diagram, from $T_{\rm m}$ to $\simeq 100 T_{\rm m}$.

\subsection{Practical formula}

Figure~\ref{Fig2} (b) and (c) present closer look on the weaker ($\Gamma/\Gamma_{\rm m}\lesssim 0.06$) and stronger  ($0.1\lesssim \Gamma/\Gamma_{\rm m}\lesssim 1$)  coupling portions of the data, respectively. Two different functions ${\mathcal F}$ can fit these data sets separately. The first fit,       
\begin{equation}\label{fit1}
{\mathcal F}_1(x)\simeq 0.104/x^{0.4},
\end{equation}
is shown in Figure~\ref{Fig2} (b). This is just an {\it ad hoc} approach for the intermediate coupling regime. There is no physical background behind this choice; the particular functional form has been chosen because of its simplicity. As a consequence, this fit does not (and is not expected to) reproduce the weak coupling asymptote~\cite{DaligaultPRE2014} $\eta_{\rm R}\propto 1/(\Gamma^2\Lambda)$ at $\Gamma\ll 1$, where $\Lambda$ is the Coulomb logarithm. 

More attention should be given to the strongly coupled regime not too far from the fluid-solid phase transition, where some theoretical predictions are available. Here, although the functional form ${\mathcal F}(x)$ can be system-dependent, certain universality has been previously suggested. For example, Vaulina {\it et al}.~\cite{VaulinaJETP2004,VaulinaAIP2005} using the activation energy ideas suggested that the shear viscosity of Yukawa fluids scales as  $\eta_{\rm R}\propto \exp(bx)$ on approaching freezing. 
They proposed to use this dependence in the regime $x\gtrsim 0.5$, where they found $b\simeq 2.9$. Similar scaling was proposed by Kaptay~\cite{Kaptay2005} for the viscosity of pure liquid metals. He arrived at this scaling combining the Andrade's equation~\cite{Andrade1931,Andrade1934} with either the activation energy concept or the free volume arguments. Testing this scaling on the experimental data for 15 selected liquid metals the adjustable parameter $b$ was obtained as $b=2.3\pm 0.2$. In a recent study by Costigliola {\it et al}.~\cite{CostigliolaJCP2018} dealing with computer simulations of viscosity in the Lennard-Jones liquid and experimental data for argon and methane this scaling has been confirmed only in the close vicinity of the melting temperature. Systematic deviations have been observed when moving away from the melting point and a different scaling of the form ${\mathcal F}(x)\propto \exp(B\sqrt{x})$ has been put forward. Both expressions are consistent with the isomorph theory (freezing temperature scaling)~\cite{CostigliolaJCP2018} and we have a good opportunity to test which of the scalings performs better in the special case of Yukawa fluids. We therefore fitted the data corresponding to the strongly coupled regime $0.1\lesssim \Gamma/\Gamma_{\rm m}\lesssim 1$ using the two functional forms. The results are shown in Fig.~\ref{Fig2} (c). The simple exponential scaling with $b=2.28$ (dashed curve) does a good job, but only at $\Gamma/\Gamma_{\rm m}\gtrsim 0.3$. The Costigliola's formula with $B=3.64$ (solid curve) allows to describe well the data in the entire range considered. Thus, this latter scaling is superior, at least for Yukawa fluids. The explicit expression for the reduced shear viscosity of strongly coupled Yukawa fluids is                      
\begin{equation}\label{fit2}
{\mathcal F}_2(x) \simeq 0.126 \exp\left(3.64 \sqrt{x}\right),
\end{equation}

We can now combine the expressions for ${\mathcal F}_1$ and ${\mathcal F}_2$ in the following form, 
\begin{equation}\label{inter}
\eta_{\rm R} = \left({\mathcal F}_1^{\gamma}+ {\mathcal F}_2^{\gamma} \right)^{1/\gamma},
\end{equation}
which then agrees with the individual expressions in their corresponding regimes of applicability and appropriately interpolates
between them. The interpolation formula (\ref{inter}) is shown in Fig.~\ref{Fig2}(a)  as the blue solid curve. The adjustable parameter $\gamma$ is set to $\gamma = 4$. The viscosity is predicted correctly within $\simeq 10\%$ tolerance as indicated by the dashed curves (only 3 from about 90 data points are clearly outside the region marked by the dashed curves).  Equation (\ref{inter}) can also be easily rewritten in the form $\eta_*(\Gamma)$. The corresponding curves are shown in Fig.~\ref{Fig1}.

\subsection{Alternative formula}

An alternative expression for the shear viscosity of Yukawa fluids can also be elaborated. Bastea proposed an accurate three-term fit to describe the behavior of the OCP fluid viscosity $\eta_{*}$ obtained in MD simulation in a wide range of coupling.~\cite{BasteaPRE2005} As we have observed in figure~\ref{Fig2}(a), the dependence of $\eta_{\rm R}$ on $x$ is practically insensitive to $\kappa$, at least for $\kappa\lesssim 3$. Rewriting the original fit $\eta_{*}(\Gamma)$ into the form $\eta_{\rm R}(x)$ we obtain the following generalization:         
\begin{equation}\label{OCPfit}
\eta_{\rm R} \simeq 0.00022 x^{-3/2}+0.096x^{-0.378}+4.68 x^{3/2}.
\end{equation}
This fit is shown by the red solid curve in Fig.~\ref{Fig2}(a). This fit can be particularly useful near the OCP limit at $\kappa\lesssim 1$.

\section{Some tendencies in the strong coupling regime}

\begin{table}
\caption{\label{Tab1} Reduced shear viscosity coefficient $\eta_{\rm R}$  and the product $D\eta(a/T)$ of several liquid metals at the corresponding melting temperatures as calculated from the data summarized in Ref.~\onlinecite{MarchBook}.  }
\begin{ruledtabular}
\begin{tabular}{lrrrrrrr}
Metal & Li & Na & K & Rb & Cu & Ag &  In   \\ \hline
$\eta_{\rm R}$ & 5.6 & 5.9 & 5.7 & 5.7 & 5.1 & 5.1 & 5.1  \\
$D\eta(a/T)$ & 0.10 & 0.12 & 0.11 & 0.12 & 0.13 & 0.10 & 0.10 \\
\end{tabular}
\end{ruledtabular}
\end{table}

At the melting point, both Eq.~(\ref{fit2}) and (\ref{OCPfit}) yield the viscosity coefficient $\eta_{\rm R}\simeq 4.8$. Quasi-universality in the dependence $\eta_{\rm R}(x)$ clearly implies quasi-universality of $\eta_{\rm R}$ values at the melting point ($x=1$). This can be rewritten as
\begin{equation}\label{AndradeScaling}
\eta_{\rm m} = C n^{2/3}\sqrt{T_{\rm m}m}, 
\end{equation}
where the subscript ``m'' stands again for the melting point and $C$ is approximately constant. This coincides with the scaling proposed by Andrade for liquid metals at the melting point.~\cite{Andrade1931,Andrade1934} However, the constant $C$ is only approximately universal even for simple systems. For soft repulsive Yukawa systems considered here the value of the constant ($C\simeq 4.8$) is smaller than that for Lennard-Jones liquid ($C\simeq 5.2$) and argon ($C\simeq 5.8$), recently reported.~\cite{CostigliolaJCP2018} Liquid metals also demonstrate somewhat higher $C$, as illustrated in Tab.~\ref{Tab1}, where the reduced shear viscosity coefficients of some liquid metals at the melting temperature have been evaluated using experimental data summarized by March and Tosi.~\cite{MarchBook} It was previously reported that the Yukawa viscosity model of liquid metals near melt predicts viscosities that are too low.~\cite{MurilloHEDP2008}      
     
The Stokes-Einstein (SE) relation between the coefficients of viscosity and diffusion can be written as
\begin{equation}\label{SE}
D=\frac{T}{6\pi \eta R},
\end{equation} 
where $D$ is the diffusion coefficient of a sphere of radius $R$ immersed in a medium characterized by the shear viscosity $\eta$. Applying the SE relation to atomistic scales (although this does not always works satisfactory, see Ref.~\onlinecite{BrilloPRL2011} and references therein), we arrive at 
\begin{equation}\label{SE1}
D\eta (a/T) = {\rm const}, 
\end{equation}
where the characteristic interparticle separation $a$ plays the role of the sphere radius $R$. Table~\ref{Tab1} demonstrates that this conditions is satisfied to a reasonable accuracy by liquid metals at the melting temperature.  Next, combine Eq.~(\ref{SE1}) with the de Gennes scaling of the self-diffusion coefficient in atomic liquids,~\cite{deGennes1959} $D\simeq v_{\rm T}^2/\Omega_{\rm E}$, where $\Omega_{\rm E}$ is the Einstein frequency (this scaling has been recently verified for single-component Yukawa fluids~\cite{KhrapakJPCO2018}). This yields
\begin{equation}
\eta \sim m\Omega_{\rm E}/a,
\end{equation}
which coincides (to within a numerical coefficient of order unity) with the expression obtained by Andrade using completely different arguments.~\cite{Andrade1931,Andrade1934} Having derived this expression he argued that ``when a solid is melted it still retains in the liquid form sufficient of its crystalline character for the molecules to possess a frequency of vibration which is practically the same as that of a solid form at the melting point''.~\cite{Andrade1931} Quantitatively, this means that the Einstein frequency is not expected to change much across the fluid-solid phase transition. Indeed, for weakly screened Yukawa systems the Einstein frequency is only slightly higher in a fluid phase as compared to an ideal crystal, as has been recently shown theoretically~\cite{KhrapakPoP2014,KhrapakPoP03_2018} and documented experimentally (using a strongly coupled dusty plasma).~\cite{WongIEEE2018} According to the Lindemann melting rule, $\Omega_{\rm E}\propto \sqrt{T_{\rm m}/m a^2}$ at the melting point, and this immediately leads to the scaling of Eq.~ (\ref{AndradeScaling}). This  route provides an alternative derivation of the Andrade' scaling (\ref{AndradeScaling}).

\begin{figure}
\includegraphics[width=10cm]{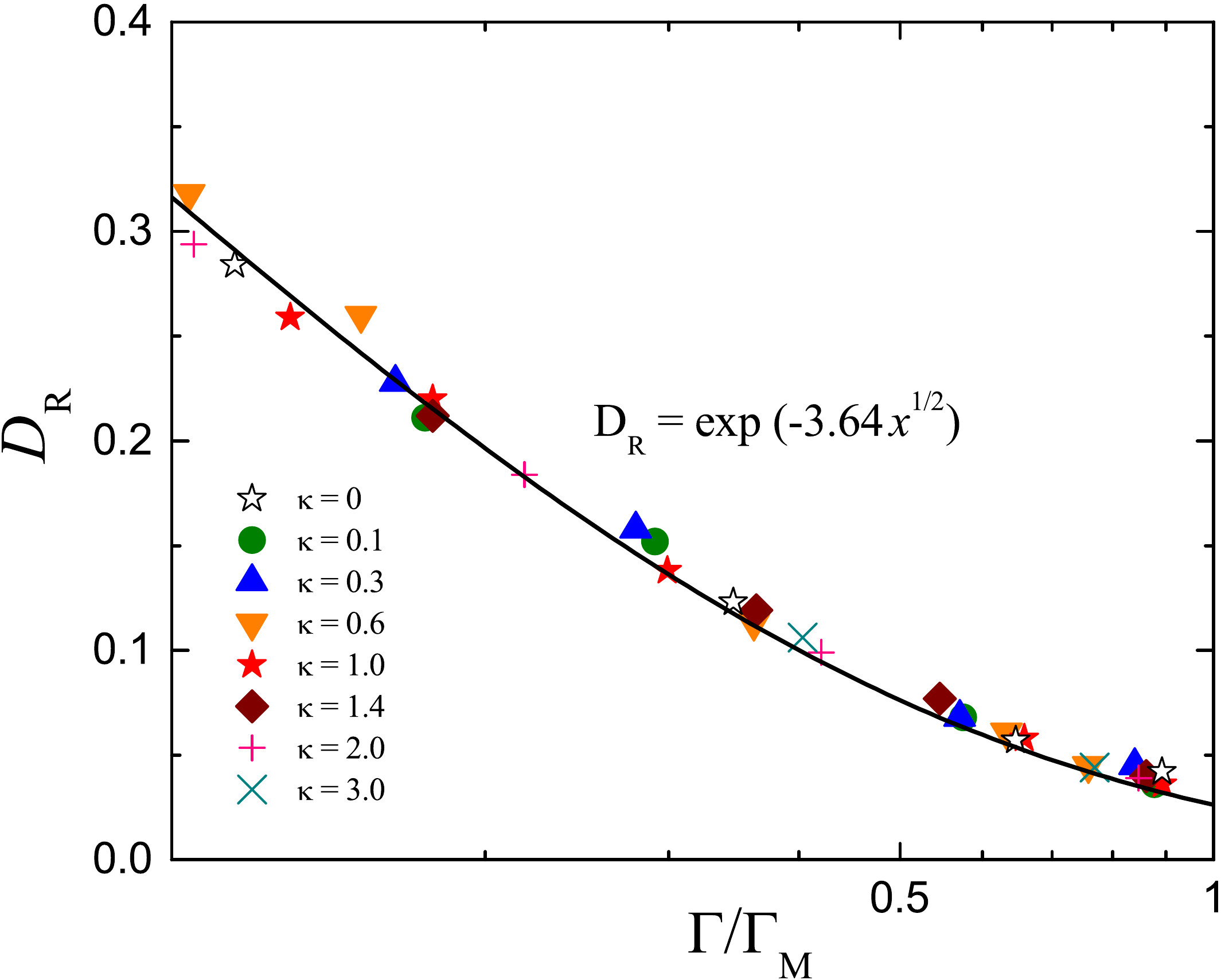}
\caption{Reduced self-diffusion coefficient $D_{\rm R}$ in Yukawa fluids as a function of the reduced coupling parameter $\Gamma/\Gamma_{\rm m}$. Numerical data are shown by symbols. The data for the OCP fluid are from Ref.~\onlinecite{HansenPRA1975}. The data for the Yukawa fluid are from Ref.~\onlinecite{OhtaPoP2000}. The solid curve corresponds to $D_{\rm R}\simeq  \exp(-3.64\sqrt{x})$, and it fits very well the numerical data. }
\label{Fig3}
\end{figure}

Another consequence of the SE relation is that the reduced self-diffusion coefficient of Yukawa fluids, $D_{\rm R}=Dn^{1/3}/v_{\rm T}$, is expected to scale as $\propto \exp(-3.64\sqrt{x})$ on approaching melting. This scaling is verified in Fig.~\ref{Fig3} using the data tabulated in Refs.~\onlinecite{HansenPRA1975,OhtaPoP2000}, which have been re-scaled to the present dimensionless form.~\cite{KhrapakPoP2012}  It is observed that the dependence $D_{\rm R}\simeq  \exp(-3.64\sqrt{x})$ describes the data quite well in the extended range of coupling. The value $D_{\rm R}\simeq 0.03$ at freezing is consistent with the values reported for several other simple model fluids at freezing (e.g. OCP, Hertzian, Gaussian-core, and inverse-power-law models).~\cite{KhrapakJPCO2018,PondSM2011,KhrapakPoP2013}  
For the strongly coupled Yukawa fluids the SE relation is of the form $D_{\rm R}\eta_{\rm R}\simeq 0.13$ and $D\eta(a/T)\simeq 0.08$. The latter value is somewhat smaller than those characterizing liquid metals at the melting temperatures (see Table~\ref{Tab1}).                   

 \section{Concluding remarks}

A simple practical expression for the shear viscosity coefficient of 3D Yukawa fluids has been put forward. The proposed formula is applicable in a wide range of coupling and screening and demonstrates reasonable accuracy. The analyzed numerical results related to shear viscosity support the temperature scaling $\eta\propto \sqrt{T}\exp(B\sqrt{T_{\rm m}/T})$ (at a constant density) on approaching the fluid-solid phase transition.~\cite{CostigliolaJCP2018} Combining this scaling with the Stokes-Einstein relation allows to reproduce quite well the behavior of the self-diffusion coefficient on approaching freezing. Note that this scaling operates in a very wide temperature (coupling) range. In the nearest vicinity of the melting point other scalings can potentially be more appropriate or accurate.              

In the context of complex (dusty) plasmas many experiments and simulations have been dealing with two-dimensional (2D) mono-layers of particles.~\cite{NosenkoPRL2004,LiuPRL2005,DonkoPRL2006,DonkoMPLB2007,FengPRE2011,FengPRE2013} In this case Yukawa (Debye-H\"uckel potential) is also considered as a reasonable first approximation for in-plane interactions. It would be interesting, therefore, to elucidate whether the freezing temperature scaling is applicable (there are favorable indications~\cite{DonkoPRL2006}) and what is the functional form of such scaling for 2D (Yukawa) systems. However, since the physics behind the transport coefficients in 3D and 2D is quite different, this topic merits a separate detailed consideration.       

\begin{acknowledgments}
I would like to thank Mierk Schwabe for a critical reading of the manuscript.
\end{acknowledgments}    

\bibliographystyle{aipnum4-1}
\bibliography{Visc_Ref}

\end{document}